\documentclass[12pt]{article}
\normalbaselines \topmargin -0.25 truein \oddsidemargin 0.30
truein \evensidemargin 0.30 truein 

\font\tbf = cmbx12

\begin{document}


\indent
\vskip 1cm
\centerline{\tbf  NON-MINIMAL COUPLING FOR
THE GRAVITATIONAL }
\vskip 0.3cm
\centerline{\tbf AND ELECTROMAGNETIC FIELDS:
A GENERAL SYSTEM OF EQUATIONS}

\vskip 0.8cm


\centerline{by}{} \vskip 0.3cm \centerline{\tbf Alexander B.
Balakin\footnote{e-mail: Alexander.Balakin@ksu.ru} } \vskip 0.3cm
\centerline{\it Kazan State University, Kremlevskaya street 18,
420008, Kazan,  Russia,} \vskip 0.5cm \centerline{and} \vskip
0.5cm \centerline{\tbf Jos\'e P. S. Lemos\footnote{e-mail:
lemos@fisica.ist.utl.pt} } \vskip 0.3cm \centerline{\it CENTRA,
Departamento de F\'{\i}sica, Instituto Superior T\'ecnico,}
\centerline{\it Universidade T\'ecnica de Lisboa}
\centerline{\it Av. Rovisco Pais 1, 1049-001 Lisboa, Portugal,}
\centerline{\it and} \centerline{\it Observat\'orio Nacional -
MCT,} \centerline{\it Rua General Jos\'e Cristino 77, 20921, Rio
de Janeiro, Brazil.}

\vskip 3cm

{\tbf Abstract} \quad {\small We
establish a new self-consistent system of equations for the
gravitational and electromagnetic fields. The procedure is based
on a non-minimal  non-linear extension of the standard
Einstein-Hilbert$-$Maxwell action. General properties of a
three-parameter family of non-minimal linear  models are
discussed. In addition, we show explicitly, that a static spherically
symmetric charged object can be described by a non-minimal model,
second order in the derivatives of the metric, when the
susceptibility tensor is proportional to the double-dual Riemann
tensor.}

\vskip 0.5cm


\newpage

\section{Introduction}

The Einstein-Maxwell theory forms the basis for other
gravitational-electromagnetic theories.  The Einstein-Maxwell
theory arises from the Einstein-Hilbert gravitational action plus
the Maxwell action. It is minimally coupled because there is no
coupling in the Lagrangian between the Maxwell part and the curvature
part. It also gives equations which are second order in
the derivatives of the metric
(as opposed to higher order), because  the Lagrangian
does not contain generic products of curvature terms (the second
derivatives of the metric that might appear in the
Einstein-Hilbert gravitational Lagrangian form a divergence of some
vector and do not contribute to give equations of
higher order). In addition
the Einstein-Maxwell theory is linear in the electrodynamics,
which means that the Maxwell Lagrangian is quadratic in
the Maxwell tensor. The property that most interest us here is the
coupling between the electromagnetic and the gravitational parts.
Thus we are led to classify gravitational-electromagnetic theories
in a useful way into two classes, according to the theory is
minimally coupled or non-minimally coupled.

The first class is minimally coupled gravitational-electromagnetism.
It can have different subclasses. One can subdivide into two
subclasses, whether the corresponding electrodynamics is linear or
non-linear. One can then also subdivide into new sublacsses according
to the gravitational action, whether the gravitational part yields a
second order theory (such as Einstein-Hilbert theory), or a higher
order theory. For instance, a minimally coupled theory, linear in
electrodynamics, and second order in the gravity part is the
standard Einstein-Maxwell theory \cite{mtw}. There are many exact
results in the framework of this theory, such as the
Reissner-Nordstr\"om solution for a charged black hole, gravity wave
solutions in electrovacuum, cosmological models with a magnetic field,
to name a few.  Another instance is a minimally coupled theory, with
non-linear electrodynamics, and second order Einstein gravity.  The
well-known models of Born and Infeld \cite{BornInf} and of Heisenberg
and Euler \cite{HeiEuler} when coupled to gravity belong to this
subclass.  One of the most interesting problems in this theory is the
search for the regular, non-singular, black holes. This search,
started by Bardeen in 1968 \cite{Bardeen}, and developed by many
authors (see, e.g., \cite{Shikin,Bronn,Burin,Berej}) led to the recent
success of Ay\'on-Beato and Garc\'ia \cite{ay1,ay2,ay3} in finding
exact solutions to the Einstein equations coupled with specific
four-parameter models of non-linear electrodynamics. And of course
there are the other cases of higher order theories coupled to Maxwell,
or to non-linear electrodynamic theories.

The second class is non-minimal coupled
gravitational-electromagnetism. It can be subdivided according to
whether the corresponding electrodynamics is linear or non-linear.
Now, in non-minimal coupled models one can no longer divide, a
priori, into second order and higher order theories, since by
definition curvature terms appear in these models which in
principle give rise in general to higher order terms in the
equations.  This second class includes non-minimal equations for
the electrodynamics containing couplings with the Riemann and
Ricci tensors and the Ricci scalar. This class is very wide and
comprises several subclasses, such as: non-minimal linear
electrodynamics plus Einstein-Hilbert term, non-minimal non-linear
electrodynamics plus Einstein-Hilbert term, non-minimal linear
electrodynamics plus Einstein-Hilbert and other pure curvature
terms, non-minimal non-linear electrodynamics plus
Einstein-Hilbert and other pure curvature terms, and others.  All
these subclasses of models belong to this second class since they
have one specific feature: the Lagrangian contains an interaction
part with specific cross-terms, including scalar products of the
Riemann tensor and its convolutions, with the Maxwell tensor.

Our goal is to study this last second class. This class is of great
interest, since the appearance of cross-terms in the Lagrangian leads
to modifications of the coefficients involving the second-order
derivatives both in the Maxwell and Einstein equations. This means, in
particular, that gravitational waves can propagate with a velocity
different from the velocity of light in vacuum, in a similar fashion
as electromagnetic waves propagate in a material medium. This new
added feature has many interesting applications in various systems and
models, such as cosmological scenarios, gravitational waves interacting
with electromagnetic fields, and charged black holes. More
specifically, in cosmology the evolution of the gravitational
perturbations may have another rate and scale. In astrophysics, the
interaction of gravitational with electromagnetic waves may lead to
time delays in the arrival of those waves, and the gravitational waves
themselves would change their own properties in a form noticeable in
gravitational wave detection.  It also leads to important
modifications of the electromagnetic and gravitational structure of a
charged black hole.

First we consider the simpler case of minimal coupling in the
electromagnetic and gravitational parts, in Section 2. Then in
Section 3 we study non-minimal coupled theories. In section 3.1 we
obtain the structure of the master equations of the non-minimal
gravitational-electromagnetic theory, for both non-linear and
linear electrodynamics.  In section 3.2 we consider in detail the
linear version of the theory. In section 3.3 we briefly study an
example.

\section{Minimal coupling of gravity and electromagnetism}

\subsection{General formalism}

In order to explain the novelty of our approach, let us first
introduce the nomenclature in the well known case of
gravitational-electromagnetic theories minimally coupled.
We will consider, generically,
high-order terms in the gravitational part,
and non-linear terms in the electromagnetic part.
The action functional is \cite{mtw}

\begin{equation}
S = \int d^4 x \sqrt{-g} \, {\cal L}_{\rm min} \,,
\label{actionminimal}
\end{equation}
where,
\begin{equation}
{\cal L}_{\rm min} = \pounds
\left[\frac{R}{\kappa}, R_{ik}R^{ik}, R_{ikmn}R^{ikmn}, ...
\right] + {\cal L}(I_{(11)}, I^2_{(12)}) \,,
\label{actionminimal2}
\end{equation}
$g$ is the determinant of the metric tensor $g_{ik}$,
and ${\cal L}_{\rm min}$ is the Lagrangian for the minimally
coupled theory. It is composed of two distinct parts, which do
not cross,
the Lagrangian $\pounds$, related to the metric field, and the
Lagrangian ${\cal L}$, related to the electromagnetic field.
The Lagrangian $\pounds\left[\frac{R}{\kappa}, R_{ik}R^{ik},
R_{ikmn}R^{ikmn}, ... \right]$ contains geometrical scalars only,
\begin{equation}
R\,,\quad R_{ik}R^{ik}\,,\quad R_{ikmn}R^{ikmn}, ...\,,
\label{geometricscalars}
\end{equation}
where $R$ is the Ricci scalar, $R_{ik}$ is the Ricci tensor,
and $R_{ikmn}$ is the Riemann tensor. The
constant $\kappa$ is equal to $\kappa = \frac{2G}{c^4}$.

The Lagrangian
${\cal L}(I_{(11)}, I^2_{(12)})$ is an arbitrary function of
the quantities $I_{(11)}$ and $I^2_{(12)}$.
$I_{(11)}$ and $I_{(12)}$ form a first set (first subscript 1)
of electromagnetic field
invariants. This first set is composed
of two invariants (denoted in the second subscript),
the first $I_{(11)}$ and the second $I_{(12)}$
invariants.
These invariants are quadratic in
the anti-symmetric Maxwell tensor $F_{ik}$, and given by
\begin{equation}
I_{(11)} \equiv \frac{1}{2} F_{ik} F^{ik} \,, \quad
I_{(12)} \equiv \frac{1}{2} F^{*}_{ik} F^{ik}\,.  \label{2}
\end{equation}
The asterisk denotes the dualization procedure, defined as follows
\begin{equation}
F^{*ik} = \frac{1}{2}\epsilon^{ikls} F_{ls} \,.  \label{3}
\end{equation}
Here $\epsilon^{ikls} = \frac{1}{\sqrt{-g}}\, {\rm e}^{ikls}$ is the
Levi-Civita tensor and
${\rm e}^{ikls}$ is the completely anti-symmetric symbol with
${\rm e}^{0123} = - {\rm e}_{0123} = 1$.
The Maxwell tensor satisfies the condition
\begin{equation}
\nabla_{k} F^{*ik} =0 \,,
\label{conditiononmaxwelltensor}
\end{equation}
where $\nabla_{k}$ is the covariant derivative. Equation
(\ref{conditiononmaxwelltensor}) can also be written as
$\nabla_i F_{kl} + \nabla_l F_{ik} + \nabla_k F_{li} = 0$.
Due to (\ref{conditiononmaxwelltensor}), the Maxwell tensor may
be represented in terms of a four-vector potential $A_i$ as
\begin{equation}
F_{ik} = \nabla_i A_{k} - \nabla_k A_{i} = \frac{\partial
A_{k}}{\partial x^i} - \frac{\partial A_{i}}{\partial x^k} \,.  \label{6}
\end{equation}
Now, the variation of the action functional (\ref{actionminimal})
with respect to the
four-vector potential $A_i$ gives the minimal vacuum Maxwell equations
\begin{equation}
\nabla_{k} \left[\frac{\partial {\cal L}}{\partial I_{(11)}} \ F^{ik} +
\frac{\partial{\cal L}}{\partial I_{(12)}} \ F^{*ik} \right] = 0 \,.
\label{equationmaxwellminimal}
\end{equation}
On the other hand, the variation of the action functional
(\ref{actionminimal}) with
respect to $g^{ik}$ yields the gravitational equations
\begin{equation}
\frac{1}{\kappa}{\rm Ein}_{ik} = T_{ik}  \,,
\label{einsteinequationsminimal}
\end{equation}
where ${\rm Ein}_{ik}$ is the corresponding non-linear
generalization of the Einstein tensor, $G_{ik}= R_{ik} -
\frac{1}{2} R g_{ik} \,$. The
tensor $T_{ik}$, defined by
\begin{equation}
T_{ik} \equiv \frac{1}{2} {\cal L} g_{ik} -
\frac{\partial{\cal L}}{\partial I_{(11)}} F_{in}F_{k}^{\cdot n} -
\frac{1}{2}\frac{\partial{\cal L}}{\partial I_{(12)}}
(F^{*}_{il}F_{k}^{\cdot l} + F_{il}F_{k}^{* l}) \,,
\label{energymomentumminimal}
\end{equation}
is the symmetric stress-energy tensor of the electromagnetic field
in vacuum. The tensor $T_{ik}$ is conserved in accordance with
the Bianchi identities
\begin{equation}
\nabla^k T_{ik} = 0  \,.  \label{11}
\end{equation}

\subsection{Example:  linear Einstein-Maxwell theory}

Before we leave this section, we give in this subsection, as a
simple example, the usual linear Einstein-Maxwell theory. It can
be obtained from equations
(\ref{actionminimal})-(\ref{energymomentumminimal}) when the
gravitational Lagrangian is given by the Einstein-Hilbert term
$\frac{R}{\kappa}$ only, and
${\cal
L}(I_{(11)}, I^2_{(12)}) \equiv I_{(11)}$.
The relations
(\ref{equationmaxwellminimal})-(\ref{energymomentumminimal})
reduce, respectively, to
\begin{equation}
\nabla_{k} F^{ik} = 0  \,,
\label{liniearmaxwell}
\end{equation}
\begin{equation}
\frac{1}{\kappa} G_{ik} = T^{(0)}_{ik}  \,,
\label{einsteinlinearmaxwell}
\end{equation}
and
\begin{equation}
\quad T^{(0)}_{ik} \equiv \frac{1}{4} g_{ik} F_{mn}F^{mn} -
F_{in}F_{k}^{\cdot n} \,, \label{linearenergymomentum}
\end{equation}
where the superscript $(0)$ denotes that the tensor $T^{(0)}_{ik}$
is the simplest part of a more general electromagnetic stress-energy
tensor. Such a formalism describes a minimal coupling of
gravitation and electromagnetism, since the right-hand-side of the
Einstein equations (\ref{einsteinequationsminimal}),
as well as the Maxwell equations
(\ref{conditiononmaxwelltensor}) and (\ref{equationmaxwellminimal})
contain metric couplings
and covariant derivatives only, while the curvature
tensor appears exclusively in the left-hand-side of
(\ref{einsteinlinearmaxwell}).

In the following section, and in contrast to the minimal
gravitational-electromag\-netic equations discussed in this
section, we consider in some detail, along the section,
a non-minimal gravitational-electromagnetic
theory both with non-linear and linear electrodynamics,
generalizing the Einstein-Maxwell theory and other minimal
theories.  This approach deals with self-consistent modifications
to both the Einstein and the Maxwell equations.

\section{Non-minimal extensions \\ of the Einstein-Maxwell Lagrangian}

\subsection{Full formalism and equations}

\subsubsection{Invariants containing the  Maxwell and dual-Maxwell tensors
coupled with the Ricci scalar  and the Riemann and Ricci
tensors}

Let us introduce the invariant scalars, quadratic in the tensors
$F^{ik}$ and $F^{*}_{ik}$ and containing the Riemann and Ricci
tensors and the Ricci scalar.  These scalars yield cross-terms,
and are the appropriate quantities for the description of
non-minimal interactions. They can be formally divided into five
sets. The first set is the trivial one, containing $I_{(11)}$ and
$I_{(12)}$ alone, as described before. The second set contains
$I_{(11)}$ and $I_{(12)}$ multiplied by $R$,
\begin{equation}
I_{(21)} \equiv \frac{R}{2} g^{im}g^{kn}F_{ik}F_{mn}  \,, \quad
I_{(22)} \equiv \frac{R}{2} g^{im}g^{kn}F^{*}_{ik}F_{mn} \,.  \label{13}
\end{equation}
The third set includes the Ricci tensor $R_{mn}$,
\begin{equation}
I_{(31)} \equiv \frac{1}{2}  R^{im}g^{kn}F_{ik}F_{mn} \,, \quad
I_{(32)} \equiv \frac{1}{2}  R^{im}g^{kn}F^{*}_{ik}F_{mn} \,.  \label{14}
\end{equation}
The fourth set is based on the convolutions of the quadratic combinations
of $F^{ik}$ and
$F^{*}_{ik}$ with the Riemann tensor
\begin{equation}
I_{(41)} \equiv \frac{1}{2}  R^{ikmn}F_{ik}F_{mn} \,, \quad
I_{(42)} \equiv \frac{1}{2} R^{ikmn}F^{*}_{ik}F_{mn} \,. \label{15}
\end{equation}
The invariants $I_{(21)}-I_{(42)}$ are chosen to be linear in the
curvature. Note also that the scalar $\frac{1}{2}
R^{im}g^{kn}F^{*}_{ik}F^{*}_{mn}$ can be reduced to a linear
combination of $I_{(21)}$ and $I_{(31)}$, and the scalar
$\frac{1}{2} R^{ikmn}F^{*}_{ik}F^{*}_{mn}$ can be represented as a
linear combination of $I_{(21)}$, $I_{(31)}$, $I_{(41)}$. Finally,
the fifth set includes the various scalars nonlinear in the
curvature. Below we introduce a few of them,
\begin{equation}
I_{(51)} \equiv \frac{1}{2} g^{im}g^{kn}F_{ik}F_{mn} f_{{\rm R}} \,, \quad
I_{(52)} \equiv \frac{1}{2} g^{im}g^{kn}F^{*}_{ik}F_{mn} F_{{\rm R}} \,,
\label{fifthsetequation1}
\end{equation}
\begin{equation}
I_{(53)} \equiv \frac{1}{2}  R^{im}R^{kn}F_{ik}F_{mn} \,, \quad
I_{(54)} \equiv \frac{1}{2}  R^{im}R^{kn}F^{*}_{ik}F_{mn} \,,  \label{17}
\end{equation}
\begin{equation}
I_{(55)} \equiv \frac{1}{2}  R^{ikab}R_{abmn}F_{ik}F^{mn} \,, \quad
I_{(56)} \equiv \frac{1}{2}  R^{ikab}R_{abmn}F^{*}_{ik}F^{mn} \,, \label{18}
\end{equation}
\begin{equation}
I_{(57)} \equiv \frac{1}{2}  R^{*ikab}R^{*}_{abmn}F_{ik}F^{mn} \,,
\quad I_{(58)} \equiv \frac{1}{2}
R^{*ikab}R^{*}_{abmn}F^{*}_{ik}F^{mn} \,, \label{19}
\end{equation}
\begin{equation}
I_{(59)}  \equiv \frac{1}{2}  R^{ikab}R_{abcd}R^{cdmn}F_{ik}F_{mn} \,, ... \,.
\label{20}
\end{equation}
In equation (\ref{fifthsetequation1}), $f_{{\rm R}}$ and $F_{{\rm
R}}$ denote arbitrary functions of the all possible independent
non-linear scalar invariants of the gravitational field, such as
$R^2$, $R_{mn}R^{mn}$, $R_{ikmn}R^{ikmn}$, $R^{*}_{ikmn}R^{ikmn}$,
...

Thus, the non-minimal Lagrangian can be written in the form
\begin{equation}
{\cal L}_{{\rm non{-}min}} {=} \pounds\left(\frac{R}{\kappa},
R_{mn}R^{mn}, ...\right) + {\cal L}(I_{(11)}, I^2_{(12)},
I_{(21)}, I^2_{(22)}, I_{(31)}, I^2_{(32)}, I_{(41)}, I^2_{(42)},
...) \,. \label{actionfunctionalnonminimal}
\end{equation}
This non-minimal Lagrangian  is $U(1)$ gauge
invariant since it contains the Maxwell
tensor $F_{ik}$ only, and does not include the potential
four-vector $A^i$.

\subsubsection{Non-minimal non-linear electrodynamics}

The variation, with respect to the 4-vector $A_k$, of the action
functional with Lagrangian (\ref{actionfunctionalnonminimal})
yields the equation for the non-minimal non-linear electromagnetic field,
\begin{equation}
\nabla_k H^{ik} = 0 \,,
\label{non-minimalnon-linearelectromagneticequation}
\end{equation}
where $H^{ik}$ is the induction tensor given by
\begin{equation}
H^{ik} = {\cal V}^{ikmn}F_{mn} + {\cal W}^{ikmn}F^{*}_{mn} \,,
\label{non-minimalnon-linearelectromagneticfield}
\end{equation}
with
\begin{eqnarray}
{\cal V}^{ikmn} &\equiv&
\frac{1}{2}\,(g^{im} g^{kn} - g^{km} g^{in})\left[
\frac{\partial {\cal L}}{\partial I_{(11)}}  +
\frac{\partial {\cal L}}{\partial I_{(21)}} R \right] +  \nonumber\\
&&+\frac{1}{4}\,(R^{im} g^{kn} - R^{in} g^{km} + R^{kn} g^{im} -
R^{km} g^{in})
\frac{\partial {\cal L}}{\partial I_{(31)}} \nonumber\\
&&+  R^{ikmn} \frac{\partial {\cal L}}{\partial I_{(41)}} + ... \,,
\label{definitonofnu}
\end{eqnarray}
and
\begin{eqnarray}
&&{\cal W}^{ikmn} \equiv \frac{1}{2}\,(g^{im} g^{kn} - g^{km}
g^{in})\left[ \frac{\partial {\cal L}}{\partial I_{(12)}} +
\frac{\partial {\cal L}}{\partial I_{(22)}} R \right] +  \nonumber\\
&&+ \frac{1}{8}
\left[R(g^{im} g^{kn} - g^{km} g^{in}) -
(R^{im} g^{kn} - R^{in} g^{km} + R^{kn} g^{im} - R^{km} g^{in}) \right]
\frac{\partial {\cal L}}{\partial I_{(32)}} \nonumber\\
&&+\left[R^{ikmn} {+} \frac{R}{4}(g^{im} g^{kn} {-} g^{km} g^{in})
-\frac{1}{2} (R^{im} g^{kn} {-} R^{in} g^{km} {+} R^{kn} g^{im} {-}
R^{km} g^{in}) \right]\frac{\partial {\cal L}}{\partial I_{(42)}} \nonumber\\
&&+ ... \,.
\label{definitonofW}
\end{eqnarray}

\subsubsection{The generalized  higher order Einstein equations}

The variation, with respect to the metric coefficients $g^{ik}$, of the
action functional with Lagrangian
(\ref{actionfunctionalnonminimal}) yields the mon-minimal extension of the Einstein
equations,
\begin{eqnarray}
0 &  =&  {\rm Ein}_{ik}
- \frac{1}{2} {\cal L} g_{ik} +
\left( \frac{\partial{\cal L}}{\partial I_{(11)}} +
R \frac{\partial{\cal L}}{\partial I_{(21)}}  \right) F_{in}F_{k}^{\cdot n} +
\nonumber\\
&&
+ \frac{1}{2}\left(\frac{\partial{\cal L}}{\partial I_{(12)}} +
R \frac{\partial{\cal L}}{\partial I_{(22)}} \right)
(F^{*}_{il}F_{k}^{\cdot l} + F^{*}_{kl}F_{i}^{\cdot l}) +
R_{ik} \left( I_{(11)} \frac{\partial{\cal L}}{\partial I_{(21)}} +
I_{(12)} \frac{\partial{\cal L}}{\partial I_{(22)}} \right)
\nonumber\\
&&
+ \left( g_{ik} \nabla^l\nabla_l - \nabla_i \nabla_k \right)
\left(I_{(11)} \frac{\partial{\cal L}}{\partial I_{(21)}} +
I_{(12)} \frac{\partial{\cal L}}{\partial I_{(22)}}\right) +
\nonumber\\
&& {+} \frac{1}{2} \frac{\partial{\cal L}}{\partial I_{(31)}}
\left[ F^{ln}(R_{il}F_{kn} {+} R_{kl}F_{in}) {+}
R^{mn}F_{im}F_{kn} \right] {+} \frac{1}{4}g_{ik} \nabla_{m}
\nabla_{l} \left(\frac{\partial{\cal L}}{\partial
I_{(31)}}F^{mn}F^{l}_{\cdot n} \right)
\nonumber\\
&& {+} \frac{1}{4} \nabla^m \nabla_m \left(\frac{\partial{\cal
L}}{\partial I_{(31)}}F_{in}F_{k}^{\cdot n} \right) {-}
\frac{1}{4}\nabla_l \left[ \nabla_i \left(\frac{\partial{\cal
L}}{\partial I_{(31)}}F_{kn}F^{ln}\right) {+} \nabla_k \left(
\frac{\partial{\cal L}}{\partial I_{(31)}} F_{in}F^{ln}\right)
\right]
\nonumber\\
&& {+} \frac{1}{8} \nabla_m \left\{\nabla^m
\left[\frac{\partial{\cal L}}{\partial I_{(32)}} (
F^{*}_{in}F_{k}^{\cdot n} {+} F_{in}F_{k}^{* n}) \right] {+}
g_{ik}  \nabla_{l} \left[\frac{\partial{\cal L}}{\partial
I_{(32)}}(F^{*mn}F^{l}_{\cdot n} {+} F^{mn}F^{*l}_{ \ \cdot
n})\right] \right\}
\nonumber\\
&& {-} \frac{1}{8}\nabla_l \left\{
\nabla_i\left[\frac{\partial{\cal L}}{\partial I_{(32)}}
(F^{*}_{kn}F^{ln} {+} F_{kn}F^{*ln})\right] {+} \nabla_k \left[
\frac{\partial{\cal L}}{\partial I_{(32)}} (F^{*}_{in}F^{ln} {+}
F_{in}F^{*ln}\right] \right\} {+}
\nonumber\\
&& + \frac{1}{16}\frac{\partial{\cal L}}{\partial I_{(32)}}\left[
R (F^{*}_{in}F_{k}^{\cdot n} + F^{*}_{kn}F_{i}^{\cdot n} +
F_{in}F_{k \cdot }^{* n} + F_{kn}F_{i \cdot}^{* n}) + \right.
\nonumber\\
&&\quad\quad\quad\quad\quad
\left. + (F^{*}_{in} R_{km} + F^{*}_{kn} R_{im})F^{mn} +
(F_{in} R_{km} + F_{kn} R_{im})F^{*mn} \right]
\nonumber\\
&&
+ \frac{3}{4} \frac{\partial{\cal L}}{\partial I_{(41)}}
F^{ls}(F_{i}^{\cdot n}R_{knls}+F_{k}^{\cdot n}R_{inls}) +
\frac{1}{2}\nabla_{m} \nabla_{n}
\left[\frac{\partial{\cal L}}{\partial I_{(41)}}
\left(F_{i}^{\cdot n}F_{k}^{\cdot m} + F_{k}^{\cdot n}F_{i}^{\cdot m}
\right)\right] +
\nonumber\\
&&
+ \frac{3}{8} \frac{\partial{\cal L}}{\partial I_{(42)}}
(F_{i}^{* m}R_{kmls} F^{ls} + F_{k}^{* m}R_{imls} F^{ls} +
F_{i}^{\cdot m}R_{kmls} F^{*ls} + F_{k}^{\cdot m}R_{imls} F^{*ls}) +
\nonumber\\
&& + \frac{1}{2}g_{ik} \frac{\partial{\cal L}}{\partial I_{(42)}}
I_{(42)}  + \frac{1}{4} \nabla_{m}
\nabla_{n}\left[\frac{\partial{\cal L}}{\partial I_{(42)}}
\left(F_{i}^{* m}F_{k}^{\cdot n} {+} F_{k}^{* m}F_{i}^{\cdot n} +
F_{i}^{\cdot m}F_{k}^{* n} {+} F_{k}^{\cdot m}F_{i}^{* n}
\right)\right]
\nonumber\\
&&
+...\, .
\label{einsteinnonminimal}
\end{eqnarray}
These equations can be rewritten in the well-known form ${\rm
Ein}_{ik} = \kappa T^{{\rm eff}}_{ik}$, but it is not the canonic
form when one is dealing with general non-minimal non-linear
electrodynamics. The reason is the following: even if the
Lagrangian for the pure gravitational field is of the
Einstein-Hilbert form, the equation (\ref{einsteinnonminimal})
contains higher order derivatives of the metric, coming from the
curvature tensor in terms containing non-minimal scalars,
$\frac{\partial{\cal L}}{\partial I_{(ab)}}$. Thus, a generic
non-minimal non-linear electrodynamics is associated with a higher
order gravitation. One can then ask the question, whether or not
non-minimal non-linear electrodynamics models exist for which
gravity is of second order. We believe that this is possible for
a special choice of the dependence ${\cal L}(I_{(ab)})$ and for
specific symmetric space-times. Below we consider a simple example
confirming such idea.

\subsection{Non-minimal coupling models, with the coupling
linear in the curvature, in Einstein-Hilbert gravity}

\subsubsection{The action}

A special case worth of discussion is when one restricts the above
theory to a Lagrangian that is Einstein-Hilbert in the gravity
term, quadratic in the Maxwell tensor and the coupling between the
electromagnetism and the metric is linear in the curvature. Thus
the theory may contain the invariants $I_{(11)}$, $I_{(21)}$,
$I_{(31)}$, $I_{(41)}$, only. Such a Lagrangian takes the form
\begin{equation}
{\cal L} = \frac{R}{\kappa} + \frac{1}{2} F_{mn}F^{mn}
+ \frac{1}{2} \,{\chi}^{ikmn} F_{ik}F_{mn} \,,
\label{simplifiednonminimal}
\end{equation}
where the quantity ${\chi}^{ikmn}$ is the susceptibility tensor.
The origin of such a terminology is the following. One obtains
from the Lagrangian (\ref{simplifiednonminimal}) with the
definition (\ref{non-minimalnon-linearelectromagneticfield}) that
the induction tensor $H^{ik}$ and the Maxwell tensor $F_{mn}$ are
linked by the linear constitutive law   (see, e.g.,
\cite{Mauginbook,HehlObukhov})
\begin{equation}
H^{ik} \equiv F^{ik} + {\chi}^{ikmn} F_{mn}  \,.
\label{inductiontensor}
\end{equation}
Another important tensor, appearing in the electrodynamics of
continuum media,  is the polarization-magnetization tensor
$M^{ik}$, defined by
\begin{equation}
4 \pi M^{ik} \equiv H^{ik} - F^{ik} \,,
\label{polarization-magnetizationtensor}
\end{equation}
and equal to
\begin{equation}
4 \pi M^{ik}  = {\chi}^{ikmn} F_{mn}  \,,
\label{1polarization-magnetizationtensor}
\end{equation}
according to (\ref{inductiontensor}). In the standard terminology
of continuum electrodynamics  \cite{Nye,landau} the
proportionality coefficients ${\chi}^{ikmn}$ form the so-called
susceptibility tensor. Generally, it has the same symmetry of the
indices transposition as the Riemann tensor, and has 21
independent components. In our case the susceptibility tensor is
linear in the curvature.

\subsubsection{Susceptibility tensor}

According to the specifications above,
the susceptibility tensor has to be of the form
\begin{equation}
{\chi}^{ikmn} \equiv  \frac{q_1 R}{2}(g^{im}g^{kn} {-} g^{in}g^{km}) {+}
\frac{q_2}{2} (R^{im}g^{kn} {-} R^{in}g^{km} {+} R^{kn}g^{im} {-}
R^{km}g^{in}) {+} q_3 R^{ikmn} \,.
\label{susceptibilitytensor1}
\end{equation}
The parameters $q_1$, $q_2$, and $q_3$ are in general arbitrary.
They have to be chosen by some ad hoc constraint, phenomenological
or otherwise. For instance, the Lagrangian of the type given by
equations (\ref{simplifiednonminimal}) and
(\ref{susceptibilitytensor1}), with $q_1=q_2=0, q_3= - \lambda$,
and $\lambda$ a constant, has been proposed phenomenologically by
Prasanna in the context of non-minimal modifications of the
electrodynamics \cite{Prasanna1,Prasanna2}. Some general
phenomenological properties of the Lagrangian
(\ref{simplifiednonminimal}) and (\ref{susceptibilitytensor1})
have been discussed by Goenner in \cite{Go}. The problem of a
phenomenological introduction of non-minimal terms into the
electrodynamic equations has been exhaustively studied by Hehl and
Obukhov \cite{hehl2}. Drummond and Hathrell \cite{drummond} have
made a qualitatively new step, they obtained modified Maxwell
equations from one-loop corrections of quantum electrodynamics in
curved spacetime. Their model is not phenomenological and
corresponds to the Lagrangian (\ref{simplifiednonminimal}) and
(\ref{susceptibilitytensor1}) with specific choices for $q_1$,
$q_2$, and $q_3$, which involve the fine structure constant and
the Compton wavelength of the electron. A quantum electrodynamics
motivation for the use of generalized Maxwell equations can also
be found, for instance, in the work of Kostelecky and Mewes
\cite{Kost1}. Accioly, Azeredo, Arag\~ao, and Mukai \cite{Accioly}
used the Prasanna electrodynamic equations to construct a special
example of a conserved non-minimal effective stress-energy tensor
containing the Riemann tensor. Exact solutions of master equations
of non-minimal electrodynamics in a non-linear gravitational wave
background were obtained and discussed in \cite{B1}-\cite{BL1}.

The susceptibility tensor ${\chi}^{ikmn}$ has the same index
symmetries as the Riemann tensor $R^{ikmn}$. Its convolutions
yield
\begin{eqnarray}
g_{kn}{\chi}^{ikmn} &=&  R^{im}(q_2 + q_3) +
\frac{1}{2}R g^{im}(3 q_1 + q_2) \,, \nonumber\\
g_{kn}g_{im}{\chi}^{ikmn}&=&  R (6 q_1 + 3 q_2 + q_3) \,.
\label{susceptconvoluted}
\end{eqnarray}
The coefficients $q_1$, $q_2$, and $q_3$ are considered to be independent
phenomenological parameters.  They introduce specific cross-terms,
which describe non-minimal interactions of the electromagnetic and
gravitational fields.  Thus,
one has a three-parametric family of non-minimal models. We
now consider three specific variants in the choice of the set $q_1$,
$q_2$ and $q_3$, and see how it influences the expression for
the susceptibility tensor.

\vskip0.5cm

\noindent{\it (a) {The susceptibility tensor is
proportional to the double-dual
Riemann tensor}}

\vskip0.2cm

The gravitational analogue of the dual
Maxwell tensor $F^{*}_{ik}$, is given by the
double-dual Riemann tensor
\begin{equation}
{\cal G}_{ikmn} \equiv  {}^{*}R_{ikmn}\,^{*} \equiv \frac{1}{4}
\epsilon_{ikab}R^{abcd} \epsilon_{cdmn} \,.
\label{31}
\end{equation}
The analogy is due to the similarity of the identity
$\nabla^n F^{*}_{in} = 0$ for the Maxwell tensor, with the
identity $\nabla^n{\cal G}_{ikmn} = 0$ for the
double-dual Riemann tensor.
The convolution of the double-dual Riemann tensor gives the Einstein tensor
\begin{equation}
g^{kn}\,{\cal G}_{ikmn} = R_{im} - \frac{1}{2}\,R \,g_{im} \,.
\label{33}
\end{equation}
Now, the double-dual Riemann tensor is given by
\begin{equation}
{\cal G}^{ikmn} \equiv  - \frac{R}{2}(g^{im}g^{kn} - g^{in}g^{km}) +
(R^{im}g^{kn} - R^{in}g^{km} + R^{kn}g^{im} - R^{km}g^{in}) - R^{ikmn} \,.
\label{34}
\end{equation}
Thus, if one imposes that the susceptibility tensor
${\chi}^{ikmn}$ is proportional to the double-dual Riemann tensor,
i.e.,
\begin{equation}
{\chi}^{ikmn} =  q \ {\cal G}^{ikmn} \,,
\label{susceptequaldoubledual}
\end{equation}
one obtains from equation (\ref{susceptibilitytensor1}) a one-parameter model
with the following
values for $q_1$, $q_2$, and $q_3$:
$q_1 = q_3 = - q$, and $q_2 =  2 q$.
This can also be written as,
\begin{equation}
q_1 + q_2 + q_3 = 0 \,, \quad 2q_1 + q_2 = 0 \,.
\label{qsdoubledual2}
\end{equation}
For this one-parameter model the non-minimal Lagrangian
(\ref{simplifiednonminimal}) can be rewritten in terms of the
Ricci scalar, the Maxwell tensor, the dual Maxwell tensor and the
standard Riemann tensor as follows,
\begin{equation}
{\cal L} = \frac{R}{\kappa} + \frac{1}{2} F_{mn}F^{mn} + \frac{q}{2} R^{ikmn}
F^{*}_{ik}F^{*}_{mn} \,. \label{37}
\end{equation}

\vskip0.5cm

\noindent{\it (b) {The susceptibility tensor is proportional to the Weyl
conformal tensor}}

\vskip0.2cm

The Weyl tensor is given by
\begin{equation}
{\cal C}^{ikmn} \equiv  R^{ikmn} + \frac{R}{6}(g^{im}g^{kn} {-}
g^{in}g^{km})
{-} \frac{1}{2} (R^{im}g^{kn} {-} R^{in}g^{km} {+} R^{kn}g^{im} {-}
R^{km}g^{in})  \,.
\label{38}
\end{equation}
It has vanishing trace, i.e., $g_{kn}{\cal C}^{ikmn}= 0$.
If one imposes that the susceptibility tensor ${\chi}^{ikmn}$
is proportional to the Weyl tensor, i.e.,
\begin{equation}
{\chi}^{ikmn} =  q \ {\cal C}^{ikmn} \,, \label{susceptequalweyl}
\end{equation}
one obtains from  equation (\ref{susceptibilitytensor1}) that
\begin{equation}
3q_1 + q_2 = 0 \,, \quad q_2 + q_3 = 0 \,.
\label{qssuceptweyl}
\end{equation}
This is also a one-parameter model for which one can easily
explicitly give the non-minimal Lagrangian
(\ref{simplifiednonminimal}).

\vskip0.5cm \noindent{\it (c) {The susceptibility tensor is equal
to the Drummond-Hathrell tensor}}

\vskip0.2cm

Drummond and
Hathrell \cite{drummond} have obtained modified Maxwell equations
from one-loop corrections in quantum electrodynamics in curved
spacetime. Their model corresponds to the Lagrangian
(\ref{simplifiednonminimal}),(\ref{susceptibilitytensor1}) with
the following coefficients
\begin{equation}
2q_1-q_3=0 \,, \quad 13q_1+q_2=0 \,, \quad q_1 = - \frac{\alpha
\lambda^2_e}{180 \pi} \,,
\label{qsdrummondhathrell}
\end{equation}
where $\alpha$ is the fine structure constant and $\lambda_{\rm
e}$ is the Compton wavelength of the electron. This is also a
one-parameter model for which one can easily explicitly give the
non-minimal Lagrangian (\ref{simplifiednonminimal}).

\subsubsection{Non-minimal constitutive equations for the
electromagnetic field}

The relation (\ref{inductiontensor}) is of the type of a linear
constitutive equation \cite{Mauginbook,HehlObukhov}
\begin{equation}
H^{ik} = C^{ikmn} F_{mn}\,,
\label{link}
\end{equation}
where the material tensor
$C^{ikmn}$ links the induction tensor with the Maxwell tensor.
Comparing (\ref{inductiontensor}) with (\ref{link}) one finds
\begin{equation}
C^{ikmn} \equiv \frac{1}{2}(g^{im}g^{kn} - g^{in}g^{km})
+ {\chi}^{ikmn}  \,. \label{materialtensor}
\end{equation}
The material tensor $C^{ikmn}$ describes the properties of the
linear response of the material to an electromagnetic field, and
contains the information about dielectric and magnetic
permeabilities, as well as about the magneto-electric coefficients
\cite{Mauginbook,landau}.

Using the medium four-velocity $U^i$, normalized such that $U^iU_i=1$,
one can decompose $C^{ikmn}$ uniquely as
\begin{eqnarray}
&C^{ikmn} = \frac12 \left(
\varepsilon^{im} U^k U^n - \varepsilon^{in} U^k U^m +
\varepsilon^{kn} U^i U^m - \varepsilon^{km} U^i U^n \right) +
\nonumber \\&
+\frac12 \left[
-\eta^{ikl}(\mu^{-1})_{ls}  \eta^{mns} +
\eta^{ikl}(U^m\nu_{l \ \cdot}^{\ n} - U^n \nu_{l \ \cdot}^{\ m}) +
\eta^{lmn}(U^i \nu_{l \ \cdot}^{\ k} - U^k \nu_{l \ \cdot}^{\ i} )
\right] \, . &  \label{44}
\end{eqnarray}
Here $\varepsilon^{im}$ is the dielectric tensor,
$(\mu^{-1})_{pq}$ is the magnetic permeability tensor,
and $\nu_{p \ \cdot}^{\ m}$ is the
magneto-electric coefficients tensor.
These quantities are defined through
\begin{eqnarray}
\varepsilon^{im} &=& 2 C^{ikmn} U_k U_n\, \nonumber\\
(\mu^{-1})_{pq}  &=& - \frac{1}{2} \eta_{pik} C^{ikmn} \eta_{mnq}\,,
\nonumber\\
\nu_{p \ \cdot}^{\ m} &=& \eta_{pik} C^{ikmn} U_n
=U_k C^{mkln} \eta_{lnp}\,.
\label{varepsilonmagneticpermeabilitymagnetoelectriccoefficients}
\end{eqnarray}
The tensors $\eta_{mnl}$ and $\eta^{ikl}$ are anti-symmetric
tensors orthogonal to $U^i$ and
defined as
\begin{equation}
\eta_{mnl} \equiv \epsilon_{mnls} U^s \,,
\quad
\eta^{ikl} \equiv \epsilon^{ikls} U_s \,.
\label{47}
\end{equation}
They are connected by the useful identity
\begin{equation}
- \eta^{ikp} \eta_{mnp} = \delta^{ikl}_{mns} U_l U^s =
\Delta^i_m \Delta^k_n - \Delta^i_n \Delta^k_m \,,
\label{usefulidentity}
\end{equation}
where the projection tensor $\Delta^{ik}$ is defined as
\begin{equation}
\Delta^{ik} = g^{ik} - U^i U^k \,.
\label{49}
\end{equation}
The generalized 6-indices $\delta-$Kronecker tensor $\delta^{ikl}_{mns}$
(see, e.g., \cite{mtw}) may be defined by a recurrent
formula through the $\delta-$Kronecker tensor with four indices,
$\delta^{ik}_{mn}$, as
\begin{equation}
\delta^{ikl}_{mns} \equiv \delta^{i}_{m}\delta^{kl}_{ns}
+\delta^{i}_{n}\delta^{kl}_{sm}
+\delta^{i}_{s}\delta^{kl}_{mn} \,, \quad
\delta^{ik}_{mn} \equiv  \delta^{i}_{m}\delta^{k}_{n} -
\delta^{i}_{n}\delta^{k}_{m} \,.
\label{6indices}
\end{equation}
Upon contraction, equation (\ref{usefulidentity})
yields another useful identity
\begin{equation}
\frac{1}{2} \eta^{ikl}  \eta_{klm} = - \delta^{il}_{ms} U_l U^s
= - \Delta^i_m \,.
\label{50}
\end{equation}
The tensors $\varepsilon_{ik}$ and $(\mu^{-1})_{ik}$ are symmetric, but
$\nu_{l \ \cdot}^{\ k}$ is in general non-symmetric. The
dot denotes the position of the second index when lowered.
These three tensors are orthogonal to  $U^i$,
\begin{equation}
\varepsilon_{ik} U^k = 0, \quad (\mu^{-1})_{ik} U^k = 0, \quad
\nu_{l \ \cdot}^{\ k} U^l = 0 = \nu_{l \ \cdot}^{\ k} U_k \,.
\label{orthog}
\end{equation}
Using the equation (\ref{materialtensor}),
one can show through straightforward
calculations that
\begin{eqnarray}
&\varepsilon^{im} = \Delta^{im} + 2 {\chi}^{ikmn} U_k U_n \,, \nonumber
\\ &
(\mu^{-1})_{pq} = \Delta_{pq} - \frac{1}{2} \eta_{pik} {\chi}^{ikmn}
\eta_{mnq} =
\Delta_{pq} - 2 \ ^{*}{\chi}^{*}_{plqs} U^l U^s \,, \nonumber
\\ &
\nu_{p \ \cdot}^{\ m} =  \eta_{pik} {\chi}^{ikmn} U_n =
- ^{*}{\chi}_{pln}^{\ \cdot \cdot \cdot \ m} U^l U^n \,,
\label{materialtensors}
\end{eqnarray}
which in turn satisfy the relations (\ref{orthog}). From the
relations given in (\ref{materialtensors}) one sees that the
non-minimal interaction of the gravitational and electromagnetic
fields effectively changes the dielectric and magnetic properties of
the vacuum, and produces a specific magnetoelectric interaction. In
this sense, under the influence of non-minimal interactions the vacuum
behaves as a material medium, called a quasi-medium.  Note from
(\ref{materialtensors}) that the tensor ${\chi}^{ikmn}$ predetermines
the changes in the dielectric properties of this quasi-medium, the
double-dual tensor $^{*}{\chi}^{*}_{plqs}$ influences its magnetic
properties, while the dual tensor $^{*}{\chi}_{pln}^{\ \cdot \cdot
\cdot \ m}$ produces magneto-electric effects.

In order to complete this analogy, one can write the relationships
between the four-vectors electric induction $D^i$ and magnetic field
$H^i$, on one hand, and the four-vectors electric field $E^i$ and the
magnetic induction $B^i$ on the other hand. These relations are
\begin{equation}
D^i = \varepsilon^{im} E_m - B^l \nu_{l \ \cdot}^{\ i} \,, \quad
H_i = \nu_{i \ \cdot}^{\ m} E_m + (\mu^{-1})_{im} B^m \,.
\label{53}
\end{equation}
The vectors $D^i$, $H^i$, $E^i$ and $B^i$ are defined by the following
formulae:
\begin{equation}
D^i = H^{ik} U_k \,, \quad
H^i = H^{*ik} U_k \,, \quad
E^i = F^{ik} U_k \,, \quad
B^i = F^{*ik} U_k \,.
\label{54}
\end{equation}
These vectors are orthogonal to the velocity four-vector $U^i$,
\begin{equation}
D^i U_i = 0 = E^i U_i \,, \quad H^i U_i = 0 = B^i U_i \,,
\label{55}
\end{equation}
and form the basis for the $F_{mn}$ and $H_{mn}$ tensors decomposition
\begin{equation}
F_{mn} = E_m U_n - E_n U_m - \eta_{mnl} B^l \,, \quad
H_{mn} = D_m U_n - D_n U_m - \eta_{mnl} H^l \,.
\label{56}
\end{equation}

\subsubsection{Master equations for the gravitational field}

We are working with a non-minimal electro-gravitational system,
with the coupling terms linear in curvature, with the additional
restrictions that the system is also linear in the Maxwell tensor,
and the gravity part is Einstein-Hilbert. In this non-minimal
theory, linear in the curvature terms, the equations for the
gravity field (\ref{einsteinnonminimal}) can be written in such a
way as to look like the standard form of Einstein equation, i.e.,
as
\begin{equation}
R_{ik} - \frac{1}{2} R \ g_{ik} = \kappa T^{({\rm eff})}_{ik} \,.
\label{standardform}
\end{equation}
The effective stress-energy tensor $T^{({\rm eff})}_{ik}$ in
the right-hand-side of (\ref{standardform}) is quad\-ratic in the
Maxwell tensor and takes the following form
\begin{equation}
T^{({\rm eff})}_{ik} = T^{(0)}_{ik} + q_1 T^{(1)}_{ik} + q_2 T^{(2)}_{ik}
+ q_3 T^{(3)}_{ik} \,.
\label{effectivestress}
\end{equation}
The linear part of the  electromagnetic stress-energy tensor
$T^{(0)}_{ik}$ is given in equation
({\ref{linearenergymomentum}). The definitions for the other three
parts of the
stress-energy tensor,
$T^{(1)}_{ik}$, $T^{(2)}_{ik}$ and $T^{(3)}_{ik}$, are
\begin{equation}
T^{(1)}_{ik} = R \ T^{(0)}_{ik} - \frac{1}{2} R_{ik} F_{mn}F^{mn} -
\frac{1}{2} g_{ik} \nabla^l \nabla_l (F_{mn}F^{mn}) +
\frac{1}{2} \nabla_{i} \nabla_{k} (F_{mn}F^{mn})  \,,
\label{part1}
\end{equation}
\begin{eqnarray}
T^{(2)}_{ik} &=&
- \frac{1}{2}g_{ik}\left[ \nabla_{m} \nabla_{l}(F^{mn}F^{l}_{\cdot n} ) -
R_{lm}F^{mn}F^{l}_{\cdot n}\right] - F^{ln}(R_{il}F_{kn} + R_{kl}F_{in}) -
\nonumber\\
&& {-} R^{mn} F_{im} F_{kn} {-} \frac{1}{2} \nabla^l
\nabla_l (F_{in}F_{k}^{\cdot n})
{+}
\frac{1}{2}\nabla_l \left[ \nabla_i(F_{kn}F^{ln}) {+} \nabla_k(F_{in}F^{ln})
\right] \,,
\label{part2}
\end{eqnarray}
\begin{equation}
T^{(3)}_{ik} =
\frac{1}{4}g_{ik} R^{mnls}F_{mn}F_{ls} {-}
\frac{3}{4}F^{ls}(F_{i}^{\cdot n}R_{knls}+F_{k}^{\cdot n}R_{inls}) {-}
\frac{1}{2}\nabla_{m} \nabla_{n}(F_{i}^{\cdot n}F_{k}^{\cdot m} {+}
F_{k}^{\cdot n}F_{i}^{\cdot m})\,.
\label{part3}
\end{equation}
Note that $T^{(3)}_{ik}$ in equation (\ref{part3}) takes the same form as the
stress-energy tensor constructed in \cite{Accioly}.

While the stress-energy tensor of the
electromagnetic field, $T^{(0)}_{ik}$, has zero trace, the
effective stress-energy tensor $T^{({\rm eff})}_{ik}$ has a
nonvanishing trace. Indeed, ${T^{({\rm eff})}} \equiv g^{ik}
T^{({\rm eff})}_{ik}$,  is given by
\begin{eqnarray}
T^{({\rm eff})}&=& - q_1 \left[ \frac{1}{2} R F_{mn}F^{mn} +
\frac{3}{2} \nabla^{k} \nabla_{k} (F_{mn}F^{mn}) \right]
\nonumber\\
&&
- q_2 \left[ R^{mn}F^{k}_{\cdot m}F_{kn} + \frac{1}{2} \nabla^k
\nabla_k (F_{mn}F^{mn}) + \nabla^{m}
\nabla_{n}(F^{kn}F_{km}) \right]
\nonumber\\
&&
- q_3 \left[ \frac{1}{2} R^{mnls}F_{mn}F_{ls} + \nabla^{m}
\nabla_{n}(F^{kn}F_{km}) \right]
\nonumber\\
&=&
- \frac{1}{2} {\chi}^{mnls}F_{mn}F_{ls} - ( q_2 + q_3) \nabla^{m}
\nabla_{n}(F^{kn}F_{km})
\nonumber\\
&&
- \frac{1}{2}(3 q_1 + q_2)\nabla^k \nabla_k (F_{mn}F^{mn}) \,.
\label{traceefective}
\end{eqnarray}
Note that the sign of the trace is not defined a priori,
depends on the specific model one uses. This feature
also happens in non-linear electrodynamic models
(see, e.g., \cite{LemosKerner}).

Equations (\ref{standardform})-(\ref{part3}) contain covariant
derivatives of the Max\-well
tensor only, and do not involve derivatives of the Riemann
tensor, Ricci tensor and Ricci scalar. Thus for a given
electromagnetic field they form a system of differential equations
containing second order partial derivatives in the metric.
Nevertheless, these equations have to be completed by the
self-consistent equations of non-minimal electrodynamics
(\ref{conditiononmaxwelltensor}),
(\ref{non-minimalnon-linearelectromagneticequation}), and
(\ref{inductiontensor}), which contain the covariant derivatives
of the Riemann tensor, Ricci tensor and Ricci scalar. In general,
the Maxwell tensor, envisaged as a solution to equations
(\ref{conditiononmaxwelltensor}),
(\ref{non-minimalnon-linearelectromagneticequation}), and
(\ref{inductiontensor}), depends on the second order partial
derivatives of the metric. Thus, in general, the equations for
the gravitational field become of fourth order. However, the
parameters $q_1$, $q_2$ and $q_3$ are arbitrary and may be fixed
in an appropriate way. So, the question of whether or not there
are models which are effectively of second order in the
derivatives of the metric is pertinent.  Below in section 3.3. we
show explicitly one such a model.

\subsubsection{Bianchi identities}

Since the Einstein tensor in the left-hand-side of equation
(\ref{standardform}) is divergence-free, the effective
stress-energy tensor (\ref{effectivestress})-(\ref{part3}) has to
be conserved, i.e.,
\begin{equation}
\nabla^k T^{({\rm eff})}_{ik} = 0 \,.
\label{bianchiI}
\end{equation}
In order to check directly that this is true, one has to use,
first, the Maxwell equations (\ref{conditiononmaxwelltensor})
and (\ref{non-minimalnon-linearelectromagneticequation})
with (\ref{inductiontensor}), and second, the Bianchi
identities and the properties of the Riemann tensor,
$\nabla_i R_{klmn} + \nabla_l R_{ikmn} + \nabla_k R_{limn} = 0$
and
$R_{klmn} + R_{mkln} + R_{lmkn} = 0\,$,
as well as the rules for the commutation of covariant derivatives,
which for vectors yields
$(\nabla_l \nabla_k - \nabla_k \nabla_l) W^i = W^m R^i_{\cdot mlk}\,$.

\subsection{An example: static spherically symmetric
gravitational and electromagnetic fields non-minimally coupled}

The line element for the static spherically symmetric model
has the form
\begin{equation}
ds^2 = B(r) \ c^2 dt^2 - A(r) \ dr^2 - r^2(d\theta^2 +
\sin^2\theta \ d\varphi^2) \,. \label{metricstatictspheric}
\end{equation}
Assume also that the electromagnetic field inherits the static
and spherical symmetries. Then the electric field potential
$A_i$ has the form $A_i = \varphi(r) \delta^0_i$. The
Maxwell tensor happens to be equal to $F_{ik}= \varphi^{\prime}
(\delta^{r}_{i}\delta^{0}_{k} - \delta^{0}_{i}\delta^{r}_{k})$,
where a prime denotes the derivative with respect to $r$.  To
characterize the electric field, it is convenient to introduce a
new scalar quantity  $E(r)$ as follows,
\begin{equation}
E^2(r) \equiv - E^i E_i =  - \frac{1}{2} F_{ik}F^{ik} =
\frac{1}{AB}F^2_{r0} = \frac{1}{AB}{\varphi^{\prime}}\,^2 \,, \label{71}
\end{equation}
where the four-vector $E^i$ is defined in equation (\ref{54}),
and the velocity
four-vector is chosen to be equal to $U^i = \delta^i_0 B^{-\frac{1}{2}}$.

To fix the sign we choose $F_{r0}= - (AB)^{\frac{1}{2}}\,E(r)$
and $F^{r0}= (AB)^{-\frac{1}{2}}\,E(r)$. For this electromagnetic
field the Maxwell equations (\ref{conditiononmaxwelltensor})
are satisfied identically, while equations
(\ref{non-minimalnon-linearelectromagneticequation}) and
(\ref{inductiontensor}) give only one non-trivial equation when
$i=0$,
\begin{equation}
\left[r^2 E(r) \left(1 + 2 \chi^{0r}_{\cdot \cdot 0r}(r) \right)
\right]^{\prime} = 0 \,. \label{72}
\end{equation}
The function $E(r)$ can then be found to be
\begin{equation}
E(r) = \frac{Q}{r^2 \varepsilon^r_r(r)}
\,, \quad {\rm where} \quad
\varepsilon^r_r(r) \equiv 1 + 2 \chi^{0r}_{\cdot \cdot 0r}(r)
\,,  \label{electricfieldspheric}
\end{equation}
and $Q$ is a constant. Assume now that the space-time with metric
(\ref{metricstatictspheric}) is asymptotically flat and
$\chi^{0r}_{\cdot \cdot 0r}(\infty) = 0$. Then, the constant $Q$
in (\ref{electricfieldspheric}) coincides with the total charge of
the object if $\varphi(r)\to \frac{Q}{r}$ at $r \to \infty$. Using
(\ref{susceptibilitytensor1}) one can compute the term
$\chi^{0r}_{\cdot \cdot 0r}(r)$,
\begin{eqnarray}
\chi^{0r}_{\cdot \cdot 0r}(r) &=& (q_1+q_2+q_3)
\left[\frac{B^{\prime \prime}}{2AB} - \frac{(B^{\prime})^2}{4
AB^2} - \frac{A^{\prime}B^{\prime}}{4 A^2 B} \right] +\nonumber\\
&&+ (2q_1 + q_2) \frac{1}{2rA} \left(\frac{B^{\prime}}{B} -
\frac{A^{\prime}}{A} \right) - q_1 \frac{1}{r^2} \left(1 -
\frac{1}{A} \right)  \,. \label{75}
\end{eqnarray}
So, from equation (\ref{electricfieldspheric}), one sees that
generally, $E(r)$ contains derivatives of the metric up to the second
order.  With such an electric field, equations
(\ref{standardform})-(\ref{part3}) for the gravitational field
become of the fourth
order. To illustrate this statement take the trace
of equation (\ref{standardform}), $R = - \kappa T^{({\rm eff})}$,
where the trace $T^{({\rm eff})}$ is given in
(\ref{traceefective}). For the metric (\ref{metricstatictspheric}) and
the electric field (\ref{electricfieldspheric})-(\ref{75}) the
trace equation takes the form
$$
\frac{1}{\kappa}\left[\frac{B^{\prime \prime}}{B} -
\frac{(B^{\prime})^2}{2
B^2} - \frac{A^{\prime}B^{\prime}}{2 A B} +
\frac{2}{r} \left( \frac{B^{\prime}}{B} - \frac{A^{\prime}}{A}\right)
- \frac{2}{r^2} \left( A - 1 \right) \right] =
$$
$$
= (E^2)^{\prime \prime} (3q_1 {+} 2q_2 {+} q_3) {+} (E^2)^{\prime}
\left[ (3q_1 {+} 2q_2 {+} q_3) \left( \frac{B^{\prime}}{2B} {-}
\frac{A^{\prime}}{2A} {+} \frac{2}{r}
\right) {+} \frac{2}{r}(q_2 {+} q_3) \right] {+}
$$
$$
+ E^2 \left[ (q_1 {+} q_2 {+} q_3) \left( {-}
\frac{B^{\prime \prime}}{B} {+}
\frac{(B^{\prime})^2}{2B^2} {+}
\frac{A^{\prime}B^{\prime}}{2 A B} + \frac{2}{r^2} \right) {-}
\right.
$$
\begin{equation}
\left.
- \frac{(2q_1 {-} q_3)}{r} \left( \frac{B^{\prime}}{B} {-}
\frac{A^{\prime}}{A}  \right)
+  \frac{2q_1 }{r^2} (A {-} 2) \right] \,.
\label{spur}
\end{equation}
Generally, equation (\ref{spur}) includes the first and the second derivatives
of the square of the electric field $E(r)$, which contains, in its turn, the
first and the second derivatives of the metric coefficients. Thus, for
generic $q_1$, $q_2$ and $q_3$ we obtain a fourth order scalar equation
for the gravity field. Direct calculations show that the equations
derived from (\ref{standardform})
for the sets of indices $tt$, $rr$, $\theta \theta$, $\varphi \varphi$
display the same features.

Now, when the susceptibility tensor is
proportional to the double-dual Riemann tensor, i.e.,
$q_1+q_2+q_3=0$ and $2q_1+q_2=0$ or $q_1=q_3=-q$ and $q_2=2\,q$,
all the derivatives disappear from the expression for
$E(r)$, providing the formula
\begin{equation}
E(r) = \frac{Q}{r^2 + 2q \left(1 - \frac{1}{A} \right)} \,.
\label{76}
\end{equation}
Thus, we recover the result obtained by M\"uller-Hoissen and Sippel in
\cite{MH} for the special model with $q_1=q_2=\gamma, q_2=-2\gamma$.
Moreover, equations (\ref{standardform})-(\ref{part3}) simplify
significantly, in particular, equation (\ref{spur}) yields
\begin{equation}
R = \frac{2 \kappa q}{r^2 A} \left[ r (E^2)^{\prime}
+ E^2 (2 - A) + \frac{r}{2} E^2 \left( \frac{B^{\prime}}{B} {-}
\frac{A^{\prime}}{A} \right) \right] \,.
\label{spur1}
\end{equation}
This equation is, evidently, of second order
with respect to the derivative ${d}/{dr}$.
For $A(\infty) = 1$ this electric field is asymptotically Coulombian.
Formally, (\ref{76}) has a form of the type discussed in
\cite{Bardeen,ay1,ay2,ay3}. We intend to consider such a model in a
future work.

\section{Conclusion}

We have established a new self-consistent system of equations
for the gravitational and electromagnetic fields. The procedure
was based on a non-minimal and non-linear extension of the standard
Einstein-Hilbert$-$Maxwell Lagrangian. The class of systems we have
studied includes non-minimal electrodynamic equations, containing
the Riemann and Ricci tensors and the Ricci scalar both in the
non-linear and linear versions.

This class of models of non-minimal and non-linear coupling of
the gravitational and electromagnetic fields
is of great interest, since the
appearance of cross-terms in the Lagrangian leads to modifications
of the coefficients involving the higher-order derivatives both in
the Maxwell and Einstein equations. This means, in particular,
that the velocity of the coupled gravito-electromagnetic waves
should differ from the speed of light in vacuum.

The general field equations obtained in the paper can in principle
be classified using the explicit dependence of ${\cal L}
(I_{(ab)})$ on $I_{(ab)}$ in the non-linear theory, whereas in the
linear theory one uses the phenomenological parameters $q_1$,
$q_2$ and $q_3$. This is important for two reasons. First, one
should search for non-minimal models in which the gravitational
field is described by equations of the second order in the
derivatives of the metric. We have shown explicitly, that static
spherically symmetric configurations satisfy such a requirement if
the susceptibility tensor is proportional to the double-dual
Riemann tensor. This model requires a detailed analysis and we
intend to consider it in a separate paper. Second, for the
non-minimal non-linear coupling between electrodynamics and
gravitation one should search for master equations (no matter
whether they are of  second or of higher order) admitting
non-singular, regular, solutions for the gravitational and
electromagnetic fields.

\section{Acknowledgments}

AB is grateful for the hospitality of CENTRA/IST in Lisbon and a grant from
the Portuguese FCT.  This work was partially funded by the Portuguese
FCT through project POCTI/FNU/44648/2002, and by the Russian RFBR
through project no 04-05-64895.

\end{document}